\documentclass{tlp}
\usepackage[dvips]{graphicx}
\usepackage{times}
\usepackage{url}
\usepackage{changebar}
\usepackage[latin1]{inputenc}

\def\manuscript{article}
\def\SP{\mbox{SICStus} \mbox{Prolog}}
\def\QP{\mbox{Quintus} \mbox{Prolog}}
\def\RDS{$\mathrm{RDS}^{\mathrm{TM}}$}

\begin{document}
\title{SICStus~Prolog---the first 25 years}
\author{Mats Carlsson and Per Mildner \\ SICS, P.O.~Box~1263, SE-164~29~Kista, Sweden}
\submitted{4th October 2009}
\revised{1th March 2010}
\accepted{22nd November 2010}

\newcommand{\caplab}[2]{\caption{\label{#1} #2}}
\nochangebars

\date{}
\maketitle
\begin{abstract}

\SP\ has evolved for nearly 25 years. This is an appropriate point in
time for revisiting the main language and design decisions, and try to
distill some lessons.  \SP\ was conceived in a context of multiple,
conflicting Prolog dialect camps and a fledgling standardization
effort.  We reflect on the impact of this effort and role model
implementations on our development.  After summarizing the development
history, we give a guided tour of the system anatomy, exposing some
designs that were not published before.  We give an overview of our
new interactive development environment, and describe a sample of key
applications.  Finally, we try to identify key good and not so good
design decisions.
\end{abstract}

\section{Introduction}

\SP\footnote{\url{http://www.sics.se/sicstus}} is a Prolog system that
has evolved for nearly 25 years.  In this \manuscript, we revisit the
factors affecting the choice of language dialects and APIs, and
summarize the more important developments that have taken place over
this time period.  We also give an in-depth description of the anatomy
of the system and its development environment.  Some key applications
are briefly described.  Several design choices that were never
published before are described herein.  We reflect on these choices,
and try to learn some lessons.

The rest of the \manuscript\ is structured as follows.
In Section~\ref{sec:dev}, we review and motivate the main phases of development.
In Section~\ref{sec:std}, we give our perspective on two important
role models for the \SP\ language, APIs and implementation: the Prolog
standardization effort, and \QP.
In Section~\ref{sec:ana}, we describe the parts of the system that are the most
interesting from a design and implementation point of view, going into
details where warranted.
In Section~\ref{sec:env}, we describe our Integrated Development
Environment (SPIDER).
In Section~\ref{sec:app}, we briefly describe some key applications.
\cbstart
Finally, we conclude with some lessons learned from the whole endeavor.
\cbend

\cbstart
\section{Development history}
\label{sec:dev}
\cbend

\SP\ is a Prolog system that ``just happened'' as opposed to being
planned in advance.  We now review the main phases of development.

\begin{description}
\item[1983] 
\cbstart
  The Warren Abstract Machine (WAM) is published and later becomes a cult tech
\cbend
  report~\cite{SRI309}, fascinating many including the first author.
\item[1985--1990] SICS is founded and recruits the first author, who
  joins the Logic Programming Systems laboratory, headed by Seif
  Haridi.  The laboratory's first and main field of research was
  or-parallel execution of Prolog.  The first author's first
  task at SICS is to develop the Prolog engine that will be the
  subject of parallelization~\cite{TOPLAS01}.  This happens in the
  informal Aurora project~\cite{NGC90} involving David H.D. Warren and
  researchers from Manchester and ANL, who provide schedulers and
  visualizers.  Subsequently another SICStus-based or-parallel effort,
  MUSE~\cite{NACLP90,IJPP94}, doing more copying and less sharing than
  Aurora, is being pursued by other SICS researchers.  At the same
  time, SICS begins distribution of \SP, which quickly becomes popular
  mainly in the academy.  Visitors Carl Kesselman and Ralph Haygood
  develop execution profilers and native code compilers, respectively.
\item[1988--1991] 
\cbstart
  A national funding agency and several companies (see the
  Acknowledgment)
\cbend
  fund the industrialization of \SP.  This provides the resources to add several pieces of
  necessary or desirable functionality, including indexed interpreted
  code, persistent term store, and multiple library modules.
\item[1991--2010] 
\cbstart
  The first author becomes fascinated by Boolean and
  finite domain constraint solvers, and such solvers appear in
  \SP~\cite{TR91:09,PLILP97}.  The \SP\ finite domain solver eventually grows into a sizable
  subsystem. More on this in Section~\ref{sec:attv}.
\cbend
\item[1994] The ISO Core Prolog standard is published, the first author
  having been an active member of the standardization committee.
  Although the standard is not perfect, contains things that would
  better have been left out, and lacks other dearly needed items, we decide
  to comply.  
\cbstart
  This leads to the release of \SP~3, a dual mode system:
  its syntax and semantics can be switched dynamically
  between ISO and pre-ISO.
\cbend
\item[1998] 
\cbstart
  Jesper Eskilson devotes his master's thesis to a
  message-passing based design of multi-threaded execution for
  \SP~\cite{PLILP98}. A prototype implementation is finished, but does
  not quite make it into a release. When Jesper leaves SICS, the
  effort runs out of steam.
\cbend
\item[1998] SICS acquires \QP\ from a UK company, which had acquired
  it from Quintus Corp.
  The reason for this move is partly economical, partly to
  get access to documentation and design choices that can be
  integrated into \SP, and partly service to the community: the
  nitty-gritty of WAM technology was not in the UK company's area of
  expertise.  SICS makes bold plans to fuse \SP\ and \QP\ into the
  Grand Unified Prolog by the year 2000.  This is not to happen, but
  the work on a successor of \SP~3 is started, influenced
  in part by the \QP\ architecture.  At the same time, \QP\ assets
  begin to make their way into the \SP~3 system.
\cbdelete
\item[2007] The shortcomings of \SP~3 and the need for a successor
  were evident since early on: in particular, its dual dialect and other
  dynamic aspects are difficult to defend and maintain; by design it
  can only use 256M of virtual memory, way too little for many
  applications.  After a major redesign, the successor version \SP~4
  is deemed ready for release.
\item[2009] 
\cbstart
  The first author finally sees the advantage of logical
  loops~\cite{Schimpf02}, and they appear in \SP~4.1.  Also,
  it has been clear for a long time that users
  have come to expect more from an integrated development environment
  than what Emacs can provide.  After a considerable implementation effort
  by the second author, we release SPIDER, our Eclipse-based IDE.
\cbend
\end{description}

\cbstart
\section{Standards and role models}
\label{sec:std}
\cbend

\SP\ was conceived in a context of multiple, conflicting Prolog
dialect camps and a fledgling standardization effort.
The first author's first encounter with a Prolog system was with
DECsystem-10 Prolog i.e.\ with the Edinburgh tradition, so there was
never any question which camp to align to.  Later, \QP\ arrived on the
scene in the same tradition, by the same lead designer, and
emerged as the de-facto standard, due to its industrial quality and
speed.  \QP\ was also among the first systems to provide designs for
features such as foreign language interface, embeddability,
customization through hook predicates and functions, and module
system.  Since \QP\ seemed to be doing everything right, it seemed
pointless to try to come up with alternative designs for these
features.  Instead, in the design of \SP, we opted for the ``imitation
is the sincerest (form) of flattery'' principle~\cite{Lacon}.

The ISO Prolog standardization effort started late, too late.  The
Prolog dialects had already diverged: basically, there were as many
dialects as there were implementations, although the Edinburgh
tradition, which had grown out of David H.D. Warren's work, was always
the dominant one.  Every vendor had already invested too much effort
and acquired too large a customer base to be prepared to make radical
changes to syntax and semantics.  Instead, every vendor would defend
his own dialect against such radical changes.  Finally, after the most
vehement opposition had been worn down in countless acrimonious
committee meetings, a compromise document that most voting countries
could live with was submitted for balloting and was approved.

Although far from perfect, we wanted to promote the standard.  At the
same time, our users had already developed vast amounts of non-compliant
code, which we had no right to break.  Our solution to this dilemma
was to provide a dual dialect system, \SP~3.

\cbstart
\section{System anatomy}
\label{sec:ana}
\cbend

This section is more or less a white paper of the current system
architecture, covering the parts of the system that are the most
interesting from a design and implementation point of view.  This
description is necessarily incomplete, and the omission of some system
component does not at all mean that its design and implementation is
trivial or uninteresting.

Before and especially after our take-over of \QP, a lot of designs and
assets have migrated into \SP, including: instruction set details,
tagging scheme, \texttt{structs} and \texttt{objects} modules, foreign
language interface, message and query systems, and memory manager.  So
in the rest of this \manuscript, we will not credit \QP\ each time.

\cbstart
\subsection{Modes of execution}
\label{sec:exe}
\cbend

Prolog code can be executed in three different modes, and each variant
comes with its pros and cons.

\begin{description}
\item[Interpreted.] Prolog clauses are stored in a form that is close
  to the source code, and are executed by an interpreter written
  either in the host language or in Prolog itself.  Such an
  interpreted is an excellent base for debuggers, and is virtually
  necessary for bootstrapping purposes even in the presence of a
  compiler.  The main disadvantage is slow execution.

\item[Native code.] Early, successful implementations such
  as~\cite{Dec10Prolog,MProlog} showed that Prolog is amenable to
  compilation to native machine code with modest to good execution
  speed. Later work~\cite{Taylor91,RoyD92} demonstrated that excellent
  execution speed can be achieved with global analysis. The main
  drawbacks of native code compilation are: the large amount of work
  that has to be invested, slow compilation, difficulty of using
  stand-alone assembler and linker tools in the compilation chain, and
  its inherent lack of portability. Also, a variant of Amdahl's
  law~\cite{Amdahl67} applies: the speedup available from compiling
  code to native code is limited by the time spent elsewhere in the
  runtime system and application code.

\item[Virtual code.] This approach can be seen as a compromise between
  the above two extremes.  Its feasibility has been demonstrated by a
  vast number of programming languages including Pascal, Forth, Lisp,
  ML, and Java. Most if not all contemporary implementations of Prolog
  use this approach, exclusively or in combination with the above two.
\end{description}

\cbstart
\subsection{Virtual machine}\label{sec:vm}
\cbend

\cbstart
\SP\ was not bootstrapped the classical way, with an interpreter
written in a host language.  First came a virtual code (WAM) compiler,
developed on another Prolog system, a WAM emulator written in C,
and a meta-interpreter.  
\cbend

The original WAM
report only treated the Horn clause subset of Prolog, so of course the
instruction set had to be enriched with instructions to support cut,
arithmetic functions, arithmetic tests, term comparison etc.  Also,
some deviations from the original WAM design were made and are
described and motivated below.  Specific
features of the \SP\ VM include the following:

\paragraph{Indexing.} In \SP, clause indexing is performed as part of
the predicate call operations (\texttt{call} and \texttt{execute}),
which index on the first argument if the callee is of the appropriate
kind.  This is done by means of a per-predicate data structure
(essentially, a hash table) that maintains an index over the
clauses. This is in contrast to the original WAM, which provides
instructions to perform such indexing.  This design decision was made
mainly for convenience of incremental compilation, which deals with
one clause at a time, but also to reduce emulator overhead.  However,
incremental compilation is by no means incompatible with having
indexing instructions; witness e.g.\ Quintus Prolog.  Furthermore,
indexable clauses use \texttt{get} instructions specialized for
matching the first argument, as shown in \cbstart
Figure~\ref{fig:get0}.  \cbend

\begin{figure*}[tbp]\centering
\framebox[\textwidth][c]{
\begin{tabular}{l|l}
\tt get\_constant\_x0 $t$    & \tt get\_nil\_x0    \\
\tt get\_structure\_x0 $f/a$ & \tt get\_list\_x0 \\
\tt get\_large\_x0 $n$ & \\
\end{tabular}
} 
\cbstart
\caplab{fig:get0}{
  Specialized \texttt{get} instructions for indexable
  clauses.  Each instruction encodes a principal functor.  The
  compiled clause for such clauses begins with one such instruction,
  instead of e.g.\ \texttt{get\_constant $t$,0}.  If the
  given clause is called with a non-variable first argument, indexing
  will kick in and only try clauses that match the given principal functor.
  Hence these instructions become no-ops, and the indexing mechanism arranges to
  skip them.  If called with a variable first argument, however, these
  instructions are not skipped and act as normal \texttt{get} instructions.
  $t$ denotes an atomic term; $n$ denotes a float or
  bignum; and $f/a$ denotes the functor of a compound term.
}
\cbend
\end{figure*}

\paragraph{Backtracking.} Taking the next alternative of a
choicepoint, and removing the choicepoint if the last alternative was
taken, is done as part of a general backtracking routine. This is
again in contrast to the original WAM, which provides instructions for
these purposes. This design decision was made for the same reasons as
for the indexing issue. However, \SP\ has retained a \texttt{try}
instruction, which creates a choicepoint if multiple clauses match a
procedure call.

\paragraph{Inlined operations.} The instruction set directly supports
primitives for cut, if-then-else, arithmetic functions and
comparisons, type tests, term comparisons, passing values to and
from foreign functions, and basic built-in predicates.

It is worth going into some detail about arithmetic, as the design
has changed quite a bit.  In \SP~3, every binary arithmetic function
had a corresponding instruction with two input and one output operand
(temporary registers) and a corresponding implementation in a C
function to dereference the inputs, compute the value depending on the
types of the inputs, and store the value (see Figure~\ref{fig:ari3}).
\cbstart
\SP~4 uses the \QP\ design, 
\cbend
which is based on two accumulators holding untagged values
throughout the evaluation of an expression, and instructions falling
into four categories, each item illustrated by the corresponding part
of Figure~\ref{fig:ari4}:

\begin{figure*}[tbp]\centering
\framebox[\textwidth][c]{
\begin{tabular}{l|l}
\tt function\_1 $f,s_1,d$     & \\
\tt function\_2 $f,s_1,s_2,d$ & \tt function\_2\_imm $f,s_1,i_2,d$ \\
\end{tabular}
} 
\cbstart
\caplab{fig:ari3}{
  \SP~3 arithmetic instructions (sample).  Every
  arithmetic function is implemented by a C function that dereferences
  and untags the inputs, computes the value depending on the types of
  the inputs, tags it, and handles any stack overflows.  The virtual
  machine merely retrieves the function to call and its inputs from the
  operands, and stores the computed value in the destination.  The right hand
  side shows the special case where a binary function takes an
  immediate second argument. $f$ denotes the C function implementing
  the instruction; $s_1$ and $s_2$ are source registers; $i_2$ is a
  source immediate value; and $d$ is the destination register.
}
\cbend
\end{figure*}

\begin{samepage}
\begin{enumerate}
\item
Loading constants and variables into one of the accumulators;
unspilling intermediate results.
\item
Applying a function to the accumulators.  The case where the operands
are integers (except bignums) is handled inline in the core emulator.
\item
Storing or unifying the value of an expression;
spilling intermediate results.
\item
Comparing the values of two expressions.
\end{enumerate}
\end{samepage}

\begin{figure*}[tbp]\centering
\framebox[\textwidth][c]{
\begin{tabular}{l|l}
\tt first\_constant $i$     & \tt later\_constant $i$     \\
\tt first\_large    $n$     & \tt later\_large    $n$     \\
\tt first\_x\_value $x$     & \tt later\_x\_value $x$     \\
\tt first\_y\_value $y$     & \tt later\_y\_value $y$     \\ \hline

\tt binop\_add            & \tt binop\_add\_imm $i$            \\
\tt binop\_subtract            & \tt binop\_subtract\_imm $i$            \\
\tt binop\_multiply         & \tt binop\_multiply\_imm $i$         \\
\tt binop\_divide            & \tt binop\_divide\_imm $i$         \\
\tt binop\_idivide          & \tt binop\_idivide\_imm $i$         \\ \hline

& \tt store\_constant $i$ \\
& \tt store\_large $n$ \\
\tt store\_x\_variable $x$     & \tt store\_x\_value $x$     \\
\tt store\_y\_variable $y$     & \tt store\_y\_value $y$     \\ \hline

\tt equal\_to $\ell$ & \tt equal\_to\_imm $i,\ell$ \\
\tt less\_than $\ell$ & \tt less\_than\_imm $i,\ell$ \\
\tt greater\_than $\ell$ & \tt greater\_than\_imm $i,\ell$ \\ 
\tt not\_equal\_to $\ell$ & \tt not\_equal\_to\_imm $i,\ell$ \\
\tt not\_less\_than $\ell$ & \tt not\_less\_than\_imm $i,\ell$ \\
\tt not\_greater\_than $\ell$ & \tt not\_greater\_than\_imm $i,\ell$ \\ 
\end{tabular}
}
\cbstart
\caplab{fig:ari4}{
\SP~4 arithmetic instructions (sample).
Let A and B denote the two arithmetic accumulators.
Top: instructions that untag and load a number into A (left) or B (right).
Second left: binary operations on A and B, leaving a value in A.
Second right: binary operations on A and an immediate operand, leaving a value in A.
Third left: instructions that tag and store the contents of A into a
Prolog variable.
Third right: instructions that compare the contents of A with a given
value, and fail if they differ.
Bottom left: instructions that compare the contents of A and B, and
branch if the comparison fails.
Bottom right: instructions that compare the contents of A and an
immediate operand, and
branch if the comparison fails.
$i$ denotes a size-limited integer constant;
$n$ denotes a float or bignum;
$x$ and $y$ denote a temporary and a permanent variable, respectively; and
$\ell$ denotes an ``else'' label.
}
\cbend
\end{figure*}

In addition, for both designs, instruction variants with immediate
operands exist, as an example of instruction merging.
Thus the \SP~4 design may seem to optimize non-trivial expressions
involving intermediate values, but with a higher setup cost due to the
initial load and final store.  Experiments have shown that the
\SP~4 design is significantly faster also on code doing only simple
integer arithmetic.  Figure~\ref{fig:arithex} shows an
example of the compilation of arithmetics.

\begin{figure*}[tbp]\centering
\framebox[\textwidth][c]{
\begin{tabular}{l}
\tt incmax(X,Y,Z) :- Z is max(X+1,Y).\\ \hline
\tt function\_2\_imm add,x(0),1,x(0)\\
\tt function\_2 max,x(0),x(1),x(0)\\
\tt unify\_value x(0),x(2)\\
\tt proceed\\ \hline
\tt first\_x\_value x(0)\\
\tt binop\_add\_imm 1\\
\tt later\_x\_value x(1)\\
\tt binop\_maximum \\
\tt store\_x\_value x(2)\\
\tt proceed \\
\end{tabular}
}
\caplab{fig:arithex}{Top: a Prolog clause containing arithmetics.
Middle: the corresponding \SP~3 VM instruction sequence.
Bottom: the corresponding \SP~4 VM instruction sequence.}
\end{figure*}

\paragraph{Conditionals.} Type and arithmetic test instructions are equipped with an
``else'' branch, which is taken if the test fails. Often, the else
branch can go to the next clause, bypassing general backtracking. This
is a ``leaner and meaner'' variant of shallow
backtracking~\cite{ICLP89} which was implemented in an early version.
These else branches somewhat complicate incremental compilation.  For
example, suppose that the first clause of predicate $P/N$ contains
such an else branch.  The compiler back-end will make it point to the
general backtracking routine.  But to enable this optimization, after
the second clause of $P/N$ has been compiled, the back-end must
revisit the else branch of the first clause and make it point to the
second clause.  Finally, the second clause must not be threaded into
the general backtracking chain of the first clause.  
\cbstart 
An example is shown in Figure~\ref{fig:else}.  
\cbend

\begin{figure*}[tbp]\centering
\cbstart
\framebox[\textwidth][c]{
\begin{tabular}{l}
\tt lifetime\_map(\_, Map) :- var(Map), !. \\
\tt lifetime\_map(DUs, Map) :-  \\
\tt ~~~~lifetime\_map(DUs, 0, Map). \\ \hline
\tt lifetime\_map/3: \\
\tt ~~~~var x(1) else L1 \\
\tt ~~~~cut \\
\tt ~~~~proceed \\
\tt L1:~get\_x\_variable x(2),x(1) \\
\tt ~~~~put\_constant 0,x(1) \\
\tt ~~~~execute lifetime\_map/3 \\
\end{tabular}
}
\caplab{fig:else}{Top: a Prolog clause containing a test allowing
to branch directly into the next clause if the test fails, bypassing
general backtracking.
Bottom: the corresponding \SP~4 VM instruction sequence.
Execution starts at the first instruction, without creating any choicepoint.}
\cbend
\end{figure*}

General disjunctions and logical loops~\cite{Schimpf02} are ``flattened'' by the
compiler into anonymous predicates. Backtracking from one disjunct to
another can use the general backtracking mechanism as well as else
branches.

\paragraph{Garbage collection support.} 
The question as to what is the best garbage collection algorithm for
Prolog is a controversial one. 
For \SP, we chose to implement a mark-and-sweep algorithm
\cite{CACM88,Sahlin91}; see also~\cite{Bevemyr94asimple} for a
detailed algorithm summary.  
As shown in~\cite{Bevemyr94asimple} and
elsewhere, mark-and-copy can run faster than mark-and-sweep,
especially if there is little live data, even if the optimization
in~\cite{Chung00} is applied.  However, there is a property that,
although not enforced by the ISO standard, a lot of existing Prolog
code relies on: preservation of variable order.  This property is
maintained by construction by mark-and-sweep, but not by
mark-and-copy.  In~\cite{Bevemyr94asimple}, several methods to cope
with this problem are listed, and they all boil down to either
disabling mark-and-copy in the presence of term comparisons or adding
extra data structures to the VM for supporting variable order.
Although we are convinced that mark-and-copy is a viable alternative
to mark-and-sweep, we found that the benefits do not outweigh the
extra complexity of having to maintain a \emph{fromspace} and a
\emph{tospace}, the extra support necessary for maintaining variable
order, the less effective memory reclamation by backtracking, and the
risk of running into unforeseen problems, what with mutables, trailed
goals, attributed variables and everything.  Last but not least, we
were guided by the ``if it ain't broke, don't fix it'' principle.

The VM handles stack overflows as follows.  At procedure
calls, if the global stack has less than a prescribed amount of free
space, it is expanded and/or garbage collected. The inlined operation
instructions also check this.  Finally, the compiler emits an
instruction to perform this test elsewhere if needed, which is rarely
the case.  We have taken the approach to ensure that all memory
reachable by the garbage collector contain valid terms. This is in
contrast to e.g.\ Quintus Prolog, which does not make such a
guarantee, and uses runtime tests to determine whether or not terms
are valid.  The main issue with ensuring validity of terms concerns
\emph{permanent variables}, which are often uninitialized at the time
garbage collection occurs. However, uninitialized locations can be
discriminated from initialized ones by scanning the VM code for past
and future operations, and this is the approach taken by \SP~4; see
Section~\ref{sec:scan}.  In \SP~3, we handled this issue by ensuring
that all permanent variables be initialized before any garbage
collection could be invoked.

As we will see later, there are several conditions that cause the
execution to be suspended at the next procedure call or inlined
operation.  The VM has a conceptual ``generic overflow flag'', which
is the disjunction of all such conditions, and a ``generic overflow
handler'', which ``pushes'' the current execution state, and then
checks and handles each condition in detail.

\paragraph{Coroutining support.} 
\cbstart
\SP\ supports goals being suspended
on attributed variables~\cite{PLILP92}. Binding an attributed
variable will set the generic overflow flag, after which the generic
overflow handler will arrange for the suspended goals to be run.  This
mechanism is described in more detail in Section~\ref{sec:attv}.
\cbend

\paragraph{Interrupt handling.} A Prolog predicate can be linked to a
UNIX signal or similar.  To ensure that the VM is in a
secure state when the interrupt is serviced, a 2-stage solution is
used: when the interrupt arrives, a primary interrupt handler sets the
generic overflow flag; and at the first opportunity, the general
overflow handler services the interrupt.

\paragraph{Floats and bignums.}  Such numbers are represented as
``boxes'' on the global stack, in a way so that they can be
distinguished from regular terms.  Certain instructions encode their
occurrences in Prolog code (see Figure~\ref{fig:large}).  As Prolog
terms, they use the same basic tag as structures, but are
distinguished by non-standard functors.

\begin{figure*}[tbp]\centering
\framebox[\textwidth][c]{
\begin{tabular}{l|l}
\tt get\_large $n,x$ & \tt get\_large\_x0 $n$ \\
\tt put\_large $n,x$ \\
\tt unify\_large $n$ \\ \hline
\tt first\_large $n$ & \tt later\_large    $n$ \\
\tt store\_large $n$ \\
\end{tabular}
}
\cbstart
\caplab{fig:large}{\SP~4 instructions encoding occurrences of floats
  and bignums.
  The top four instructions encode unification
  with such numbers.  The bottom three encode arithmetic with such numbers.
  $n$ denotes a float or bignum; $x$ denotes a temporary register.}
\cbend
\end{figure*}

\paragraph{Profiling support.}  Profiling in \SP\ is done by
instrumenting the virtual code with \emph{counter} instructions.  When
executed, such instructions simply increment a private counter.  After
execution of a benchmark, the relevant counter values are easily
gathered by scanning the virtual code.  This scheme was described in 
\cite{SLP87} and was first prototyped on an early \SP\ version.
The instrumentation is done at compile time, but could have been done
directly on existing virtual code.

Although this scheme provides exact information about the number of
predicate calls and backtracks, it cannot know exactly how much time
is spent where in the code.  To overcome this obvious limitation, one
would have to monitor the VM program counter using clock
interrupts, like \texttt{gprof}.

Another current limitation is that no call graph is maintained.  It is
often of interest to know not only how many times a predicate was
called, but also where it was called from.  Such information could
be readily provided by a small piece of extra profiling, since at
every predicate call operations (\texttt{call} and \texttt{execute}),
the VM stores the caller location in a register, for use by the
source-linked debugger.

\paragraph{Low-level considerations.}  The layout of the VM code was
partly designed, partly evolved, to minimize emulator overhead.
Pointers and constants are word aligned, but instructions are half-word
aligned, which implies that instructions that contain a pointer or
constant need to exist in a (word) \emph{aligned} and an
\emph{unaligned} variant, where one of the two variants includes a
padding half-word.  Operands denoting registers are encoded with
offsets off the base address of a register bank
as opposed to just integers.  The instruction dispatch
loop makes use of \texttt{gcc}'s computed goto extension: the instruction
opcode is encoded as an offset into a table of labels.  The table has
one \emph{read mode} and one \emph{write mode} entry per
instruction. Thus, to select mode, one just adds an offset to the
opcode.  On 64-bit platforms, instructions and their fields are twice
their size on 32-bit platforms, except operands encoding bignums and
floats.

\paragraph{Instruction merging and specialization.}  These are two
well-known transformations of VM instruction sets, aiming at saving
time as well as space.  In~\cite{PPDP01}, we performed an extensive study of
these two transformations and their impact on the \SP\ VM. The
current instruction set was finalized based on that study. Briefly, we
use specialization to a very limited extent, only for the special
first argument \texttt{get} instructions mentioned above, and for
frequent instructions that move a value from one virtual register to
another. Merging, on the other hand, was found to pay off more and is
used extensively.  Instruction pairs as well as patterns involving
longer sequences are subject to merging.

\paragraph{Tagging schemes.}  All Prolog implementations need to
use some means of run-time typing of its terms.  Most implementations,
including \SP, use tagged pointers, i.e.\ machine addresses with a few
bits or even an extra word replaced by a bit-field that denotes the
type of term pointed to, but tagged object implementations also exist,
\cbstart
e.g.~\cite{BinProlog,OpenProlog}.  
\cbend
\SP~3 reserved the four most significant bits,
with the rationale that fewer bits would not suffice for encoding the
basic types, including bignums, floats, attributed variables, etc.
The implementation settled on using nine different tags.
Moreover, the two least significant bits
were reserved for use by the garbage collector.  The main disadvantage
of this choice was that it limited the address range of non-atomic
terms to 256M on 32-bit platforms, which is much too little for many
applications.  \SP~4, and the original WAM report, 
instead reserve the two least significant bits, plus a third bit when
the pointer is not a machine address (integers and atoms).  With this
design, no address space problems arise.  Bignums and floats use the
same tag as structures, but are distinguished by non-standard functors.
All types of variables use the same tag.  The garbage collector still
needs to store two bits for every word, so the question is, where?
The \SP~4 solution is to reserve a small part of each Prolog
stack as a bit array for use by the garbage collector.

\cbstart
\subsection{A note on code scanning}
\label{sec:scan}
\cbend

One of the advantages of VMs is the ease with which
various information can be extracted from the virtual code, usually in
time linear in the length of the code.  
\cbstart
This is for example the case
for the use-definition analysis~\cite{ASU86} that the garbage collector performs.
\SP~4 uses this technique in the following contexts:
\begin{itemize}
\item As mentioned before, test instructions are equipped with an ``else'' branch, which is
  taken if the test fails.  The compiler
  back-end must scan code containing such ``else''
  branches, making them point to the next clause.
\item The garbage collector needs to identify uninitialized local stack
  locations.  It also needs to know which temporary registers are
  live, if a global stack overflow occurred in the middle of VM
  code.  Code scanning solves both of these tasks.
\item \SP~supports a binary file format for precompiled code.  When
  creating such files, VM code and other pieces of the memory image
  are dumped, together with relocation information. Code scanning is
  used to find what relocation information to write to the file.  When
  loading such files, the VM code is not only scanned but also relocated.
  Relocatable information includes pointers to predicates, atom
  numbers, and endianness.
\item 
\cbstart
  All Prologs that the authors are aware store atoms in a table for purposes
  of representation sharing and $O(1)$ time identity test.
\cbend
  Since the table can fill up, many Prologs provide an atom garbage
  collector, which disposes of atoms that are no longer in use
  anywhere. The atom garbage collector needs to scan all relevant
  memory areas, including the VM code, to discover which atoms are
  still in use.
\item As mentioned before, \SP~provides a counter-based execution profiler.  If told to
  instrument code for profiling, the compiler inserts special counter
  instructions at certain places in the VM code.  The profiler later
  uses code scanning to reset those counters prior to profiling and to
  gather their values afterwards.
\item If an arithmetic instruction encounters an invalid argument at
  runtime, for example an atom, an error exception is raised.  By
  scanning the code around the program location, one can reconstruct a
  goal that is semantically if not syntactically identical to the
  source code where the error occurred. The decompiled goal is part of
  the error exception.
\end{itemize}
\cbend

\cbdelete
\cbstart
\subsection{Native code}
\cbend

Native code compilation for \SP\ has a long history. Starting in the
'80s we developed compilers from WAM code to Motorola 68K and SPARC.
We used a fixed mapping of WAM registers to machine registers, and
took care to seamlessly integrate all three execution modes:
\begin{itemize}
\item 
Native code calling non-native code and vice versa.
\item
Native code returning to non-native code and vice versa.
\item
Native code backtracking into non-native code and vice versa.
\end{itemize}

The compilation was not a mere macro expansion of the WAM
instructions.  In particular, read and write mode instruction streams
for compound term unification were kept separate and reasonably
optimized.  The target code was rich in calls to runtime routines, but
operations like dereferencing, \texttt{allocate}, \texttt{deallocate},
stack trimming, and write mode unification were inline.  Speedups by a
factor of 3 over virtual code were not uncommon.

\begin{sloppypar}
Later, Clark Haygood overhauled the native code compilers, the main
inventions being the intermediate languages \emph{SICStus Abstract
  Machine} (SAM) and \emph{RISCified SAM} (RISS)~\cite{ICLP94}.  SAM
was not only an intermediate language; it was also a macro assembly
language for the native code runtime kernel, containing all the
runtime routines.  He also added a MIPS back-end.  The compilation
paths from Prolog code resp.\ the runtime kernel to binary code are
shown in Figure~\ref{fig:nc}.
\end{sloppypar}

\begin{figure*}[tbp]\centering
\cbstart
\begin{tabular}{c}
\framebox[\textwidth][c]{\includegraphics[width=\textwidth]{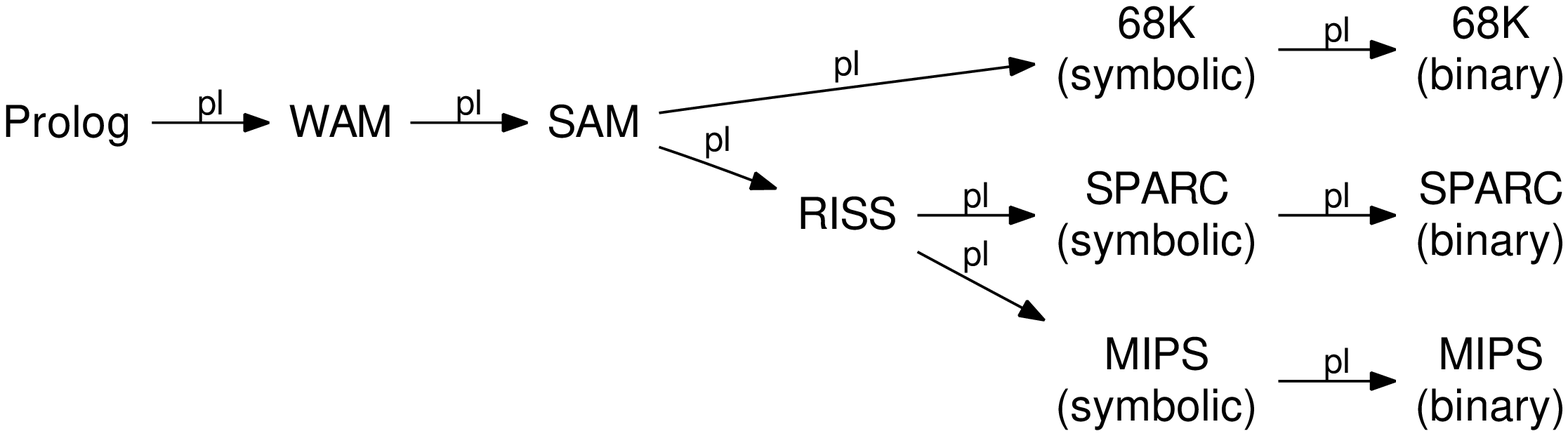}} \\ 
\ \\
\framebox[\textwidth][c]{\includegraphics[width=\textwidth]{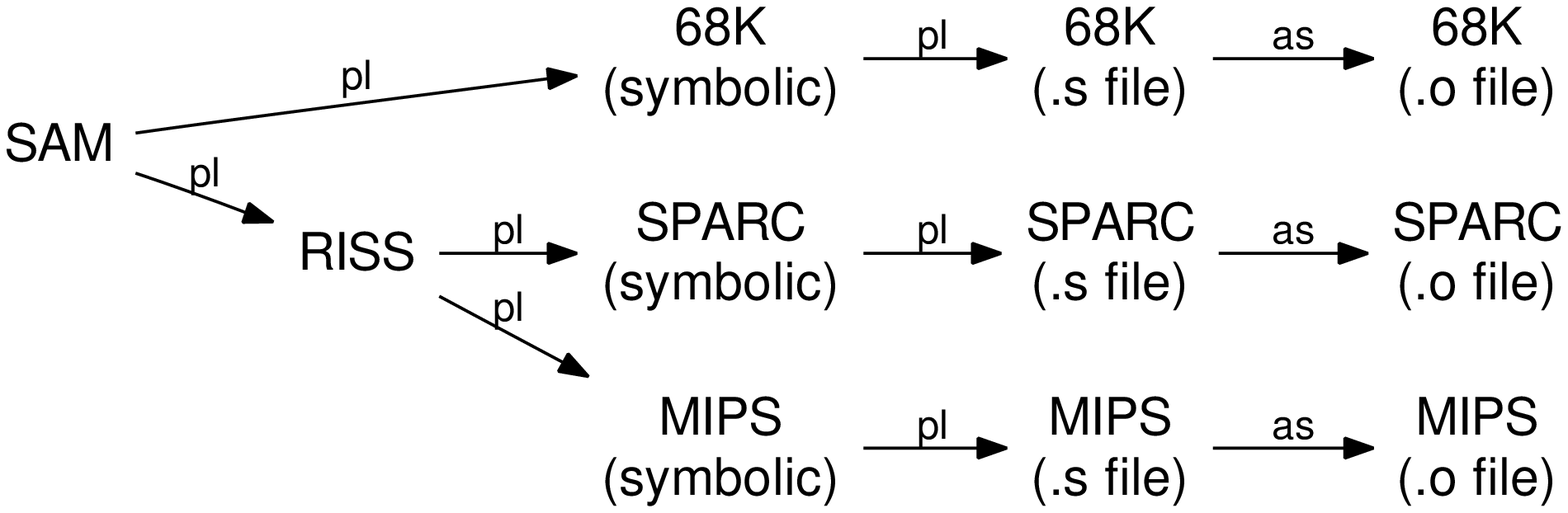}} \\ 
\end{tabular}
\cbend
\caplab{fig:nc}{Top: native code compilation path for Prolog code.
Bottom: compilation path for the native code kernel.
The standard assembler \texttt{as} is used in the native code kernel
compilation path.  Everywhere else, Prolog with the appropriate
back-ends in C is used.}
\end{figure*}

Eventually, the M68K and MIPS back-ends were dropped.  The current
\SP~3 release only supports the SPARC back-end.  Native code was
completely dropped in \SP~4 for lots of reasons, including:

\begin{itemize}
\item Amdahl's law, which tends to dominate as applications scale up.
\item The inevitably large number of wheels that tend to get
  reinvented: assembler functionality, instruction scheduling,
  register allocation, etc.
\item The difficulty of saving relocatable code in binary files and doing
  the relocation upon loading such files.
\item Scanning native code for information listed in
  Section~\ref{sec:scan} is extremely cumbersome.
\item The instruction cache easily gets confused if
  native code is modified on the fly.
\item When an architecture goes extinct, a huge investment in 
  code development is lost.
\end{itemize}

Of course, the potential of getting significant speed-up of
time-critical code is a baby that should not be thrown out with the
bathwater. JIT compilation is a well-known scheme that avoids most of
the above problems, 
\cbstart
and has been used for Prolog~\cite{YAP2007}.
\cbend
We may well explore this approach in the future.

\cbstart
\subsection{Managing dynamic code}
\cbend

Prolog makes a difference between \emph{dynamic} predicates, whose
clauses may be asserted, retracted or inspected by the running program, and
\emph{static} predicates, where such operations are not allowed.  In
practice, dynamic predicates will be represented as interpreted in the
sense of Section~\ref{sec:exe}, since accessing and inspecting clauses
is a central operation of the interpreter.  There are several
issues with interpreted and/or dynamic clauses.  We now describe how
we deal with them.

\paragraph{Indexing.} \SP\ uses the scheme for
indexing of dynamic clauses on the first argument in linear
space that was described in~\cite{DemoenMC89}.

\paragraph{Semantics.} The paper~\cite{ICLP87} proposed,
and the ISO Prolog standard later confirmed, a semantics for dynamic
clauses that are asserted or retracted during execution. The authors also invented
a clever mechanism that allows to implement the semantics in almost
constant time. The mechanism is based on a global clock register, two
time-stamps per dynamic clause, and a time-stamp per dynamic
choicepoint.  Note that a retracted clause cannot in general be
physically removed right away, as it might be in the scope of some
dynamic choicepoint.

\paragraph{Dead clause reclamation.}  It is only safe to physically
reclaim a clause when it is dead wrt.\ the global clock as well as all
dynamic choicepoints.  It would be logically correct to leave them
around, but that would of course have a disastrous effect on
performance. It is a non-trivial problem how to efficiently detect them
and organize their reclamation.  In~\cite{ICLP87}, the authors
describe how to scan for and reclaim clauses in time linear in the
number of the retracted clauses plus the number of choicepoints, but
the question is when to do it.  If it's done too often, the
choicepoint stack will be scanned over and over again for nothing.  If
it's done too seldom, dead clauses accrete, degrading performance of
dynamic code accesses.  Our implementation is a variant of this
scheme.  To make it really work, we also found it necessary:

\begin{itemize}
\item to register retracted clauses in some data structure, so that they
  can be found in $O(1)$ time,
\item to recognize and speed up the case where there are \emph{no}
  dynamic choicepoints, and
\item to recognize cases when a retracted clause can be reclaimed immediately.
\end{itemize}

\begin{sloppypar}
\paragraph{Clause references.} Although not in the ISO standard, many
Prologs provide a way to directly access a dynamic clause with a term
known as a \emph{db\_reference}.  This is provided by at least Ciao,
Quintus, SICStus, SWI and Yap Prologs.
In \SP, a db\_reference has the form
\texttt{'\$ref'($i$,$j$)} where $i$ is an integer denoting the address
of the clause, and $j$ is an integer for validation purposes; see
below.  The built-in predicate
\texttt{instance(\textit{+Ref},\textit{-Clause})} will take a
db\_reference and unify \textit{Clause} with a brand new copy of the
clause referred to. The built-in predicate
\texttt{erase(\textit{+Ref})} will retract the clause. And so on.
This feature however suffers from a dangling pointer problem.  What to
do if the clause has already been retracted? What if its memory has
been reclaimed?  We now outline how we address this problem.
\end{sloppypar}

\begin{itemize}
\item 
We maintain a global counter of asserted clauses and a global hash table that
maps the address of a clause, $i$, to the value that the counter had when
the clause was created, $j$.
\item
Db\_references are validated by checking that the hash table still
maps $i$ to $j$.
\item
Hash table entries are removed when the corresponding clause is reclaimed.
\end{itemize}

This scheme ensures that db\_references are unique, even if the memory
used by one clause happens to be reused later by another one.

\cbstart
\subsection{General memory management}
\cbend

The Prolog runtime system needs to dynamically allocate and free a
huge amount of memory blocks of sizes varying from a few bytes to
potentially several gigabytes.  The natural choice would be to use the
POSIX primitives \texttt{malloc()} and \texttt{free()}, and if code
development had started today, that would have been the likely choice.
But in the 80's, the quality of their implementations left much to be
desired.  Worse, the quality and performance varied dramatically from
platform to platform.  Also, \SP~3's requirement that certain memory
areas be allocated in a certain region of the address space is
incompatible with the standard  \texttt{malloc()} and \texttt{free()}.
So for historical and other reasons, \SP\ has its own memory manager,
the main features of which are the following:

\begin{itemize}
\item A two-layer architecture.  The bottom layer requests memory from
  the operating system (O/S) and returns memory to it.  Such requests are relatively
  infrequent and deal with \emph{bigmems}, i.e.\ relatively large
  chunks of memory.  The behavior of the bottom layer is subject to
  several tunables that the user can set.  The top layer is the
  runtime system interface.  It chops up the \emph{bigmems} into
  smaller \emph{mems}, and keeps tracks of all free \emph{mems}.
\item 
  When in use, a \emph{mem} has no header or other memory overhead.
\item The top layer keeps free \emph{mems} in multiple unsorted chains,
  each chain corresponding to a specific range of sizes.  This allows
  \emph{mems} to be allocated in almost constant time.
\item \emph{Mems} are freed in constant time---no attempt is made to
  eagerly congeal adjacent free \emph{mems}.
\item From time to time, an $O(n \log n)$ algorithm to congeal all
  adjacent free \emph{mems} is run, where $n$ is the number of free
  \emph{mems}.
\item The built-in predicate \texttt{trimcore} orders the bottom layer
  to endeavor to return \emph{bigmems} that are totally unused to the
  O/S.
\end{itemize}

The Prolog stacks tend to be the largest memory blocks by a wide
margin.  So the question arises, should a Prolog stack correspond to a
\emph{mem} or a \emph{bigmem}? It was found that treating Prolog
stacks as \emph{mems} could cause severe memory fragmentation, so our
current policy is to reserve a \emph{bigmem} for each Prolog stack.

\cbstart
\subsection{Interfacing foreign code}
\cbend

\SP\ provides multiple interfaces for calling foreign code and vice
versa.  This is not the place to describe them all, but a few points
are worth mentioning, 
\cbstart
in particular the fact that none of them exposes
the internal Prolog data structures to the foreign code.
\cbend
A comparison of such interfaces for several
implementations of Prolog can be found in~\cite{BagnaraC02TRa}.

\paragraph{Prolog-to-C interface.} The interface provides a linking of
Prolog predicates to C functions, which can succeed, fail, and raise
exceptions.  The interface does not allow to define non-deterministic
predicates.  The mapping of predicate and function names, as well as
type conversions, is declared in Prolog facts.

In \SP~3, a piece of C code is compiled from such facts for each such
procedure. This piece of code implements all necessary checks and
conversions on input arguments, calls the target functions, and
converts and unifies the output arguments as necessary.  Such code
tends to have large chunks in common from one predicate to another.

In \SP~4, the VM has instructions for such checks and conversions (see
Figure~\ref{fig:foreign}).  Foreign predicates are compiled to VM code
instead of C code.  This avoids the need to use a C compiler and
allows more code to be shared.  The only difficulty is the actual call
to the foreign function, which expects its arguments to be passed in a
way compliant with the platform ABI.  In the presence of
floating-point arguments, all call patterns cannot be precoded in
the VM emulator.  The \texttt{call\_foreign} instruction,
whose job it is to do this call, is the only part of the system that
is implemented in assembly code.  Figure~\ref{fig:foreignex} shows an
example of this compilation.

\begin{figure*}[tbp]\centering
\cbstart
\framebox[\textwidth][c]{
\begin{tabular}{l|l}
\bf Instruction & \bf TYPE \\
\tt push\_{\it TYPE} $y$ & \tt float \\
\tt push\_result\_{\it TYPE} & \tt integer \\
\tt receive\_{\it TYPE} $y$ & \tt term \\
\tt pop\_{\it TYPE} $y$ & \tt atom \\
& \tt string \\
& \tt codes \\ \hline
\tt open\_foreign\_call $\ldots$ \\
\tt call\_foreign $f,a$ \\
\tt close\_foreign\_call \\
\end{tabular}
}
\caplab{fig:foreign}{The \SP~4 instruction set for the Prolog-to-C interface.
Top left: instructions for arguments and return values.
\texttt{push\_{\it TYPE}} $y$ (for input arguments) and
\texttt{push\_result\_{\it TYPE}} (for output arguments)
populate the call frame.
\texttt{pop\_{\it TYPE}} $y$ receives an output argument.
\texttt{receive\_{\it TYPE}} $y$ receives the return value.
Top right: the types handled by the API.
Bottom left: instructions to manage the actual call.
\texttt{open\_foreign\_call} allocates the call frame,
\texttt{call\_foreign} executes the call, and
\texttt{close\_foreign\_call} deallocates the call frame.
$y$ denotes a permanent variable; $f$ is the address of the foreign
function, and $a$ is the arity.}
\cbend
\end{figure*}

\begin{figure*}[tbp]\centering
\framebox[\textwidth][c]{
\begin{tabular}{l}
\tt extern long \\
\tt ixkeys(SP\_term\_ref spec, SP\_term\_ref term, SP\_term\_ref list);\\ \hline
\tt foreign(ixkeys, c\_index\_keys(+term, +term, -term, [-integer])).\\ \hline
\tt open\_foreign\_call 4,3,c\_index\_keys/4,0\\
\tt push\_term y(0)\\
\tt push\_term y(1)\\
\tt push\_result\_term \\
\tt call\_foreign ixkeys,4\\
\tt pop\_term y(2)\\
\tt receive\_integer y(3)\\
\tt close\_foreign\_call \\
\end{tabular}
}
\cbstart
\caplab{fig:foreignex}{
Prolog-to-C interface example:
binding the predicate \texttt{c\_index\_keys/4} to the \texttt{ixkeys()} function.
Top: the header of the C function \texttt{ixkeys}.
The type \emph{SP\_term\_ref} provides a safe reference from C to a Prolog term.
Middle: the foreign declaration, from which the VM instruction sequence is generated.
Bottom: the \SP~4 VM instruction sequence for \texttt{c\_index\_keys/4}:
the first four instructions allocate and populate a call frame,
one instruction executes the call,
two instructions receive the output argument and the function value, and
the last instruction deallocates the call frame.}
\cbend
\end{figure*}

The basic interface handles simple C types only.  In addition, the
\texttt{structs} module provides a way to declare C structs in Prolog
with name-based access to their fields, and to pass struct pointers to
C code (see Figure~\ref{fig:strex}).  The \texttt{objects} module is
built on top of this feature.

\begin{figure*}[tbp]\centering
\framebox[\textwidth][c]{
\scriptsize
\begin{tabular}{l|l}
\tt :-~foreign\_type & \\
\tt ~~~~intgr~~~~=~integer\_32, & \tt typedef~int~intgr;\\
\tt ~~~~bool~~~~~=~enum([ & \tt typedef~enum~\_bool~\{\\
\tt ~~~~~~~~~~~~~~~~~~~~~~~~~~false, & \tt ~~~~~~~~false,\\
\tt ~~~~~~~~~~~~~~~~~~~~~~~~~~true & \tt ~~~~~~~~true\\
\tt ~~~~~~~~~~~~~~~~~~~~]), & \tt \}~bool;\\
& \tt typedef~struct~\_position~position;\\
\tt ~~~~position~=~struct([~~~ & \tt struct~\_position~\{\\
\tt ~~~~~~~~~~~~~~~~~~~~~~~~~~x:integer\_32, & \tt ~~~~~~~~int~x;\\
\tt ~~~~~~~~~~~~~~~~~~~~~~~~~~y:integer\_32 & \tt ~~~~~~~~int~y;\\
\tt ~~~~~~~~~~~~~~~~~~~~~~]), & \tt \};\\
& \tt typedef~struct~\_size~size;\\
\tt ~~~~size~~~~~=~struct([~~~ & \tt struct~\_size~\{\\
\tt ~~~~~~~~~~~~~~~~~~~~~~~~~~width:integer\_16, & \tt ~~~~~~~~short~width;\\
\tt ~~~~~~~~~~~~~~~~~~~~~~~~~~height:integer\_16 & \tt ~~~~~~~~short~height;\\
\tt ~~~~~~~~~~~~~~~~~~~~~~]), & \tt \};\\
& \tt typedef~struct~\_mongo~mongo;\\
\tt ~~~~mongo~~~~=~struct([~~~ & \tt struct~\_mongo~\{\\
\tt ~~~~~~~~~~~~~~~~~~~~~~~~~~a:intgr, & \tt ~~~~~~~~intgr~a;\\
\tt ~~~~~~~~~~~~~~~~~~~~~~~~~~b:integer\_16, & \tt ~~~~~~~~short~b;\\
\tt ~~~~~~~~~~~~~~~~~~~~~~~~~~c:integer\_8, & \tt ~~~~~~~~char~c;\\
\tt ~~~~~~~~~~~~~~~~~~~~~~~~~~d:unsigned\_16, & \tt ~~~~~~~~unsigned~short~d;\\
\tt ~~~~~~~~~~~~~~~~~~~~~~~~~~e:unsigned\_8, & \tt ~~~~~~~~unsigned~char~e;\\
\tt ~~~~~~~~~~~~~~~~~~~~~~~~~~f:float\_32, & \tt ~~~~~~~~float~f;\\
\tt ~~~~~~~~~~~~~~~~~~~~~~~~~~g:float, & \tt ~~~~~~~~double~g;\\
\tt ~~~~~~~~~~~~~~~~~~~~~~~~~~h:atom, & \tt ~~~~~~~~SP\_atom~h;\\
\tt ~~~~~~~~~~~~~~~~~~~~~~~~~~i:string, & \tt ~~~~~~~~char~*i;\\
\tt ~~~~~~~~~~~~~~~~~~~~~~~~~~j:address, & \tt ~~~~~~~~void~*j;\\
\tt ~~~~~~~~~~~~~~~~~~~~~~~~~~k:array(81,integer\_8), & \tt ~~~~~~~~char~(k)[81];\\
\tt ~~~~~~~~~~~~~~~~~~~~~~~~~~l:size, & \tt ~~~~~~~~size~l;\\
\tt ~~~~~~~~~~~~~~~~~~~~~~~~~~m:pointer(position), & \tt ~~~~~~~~position~*(m);\\
\tt ~~~~~~~~~~~~~~~~~~~~~~~~~~n:pointer(belch), & \tt ~~~~~~~~belch~*(n);\\
\tt ~~~~~~~~~~~~~~~~~~~~~~~~~~o:bool, & \tt ~~~~~~~~bool~o;\\
\tt ~~~~~~~~~~~~~~~~~~~~~~~~~~p:integer, & \tt ~~~~~~~~long~p;\\
\tt ~~~~~~~~~~~~~~~~~~~~~~~~~~q:pointer(mongo) & \tt ~~~~~~~~mongo~*(q);\\
\tt ~~~~~~~~~~~~~~~~~~~~~~]), & \tt \};\\
& \tt typedef~union~\_uex~uex;\\
\tt ~~~~uex~~~~~~=~union([~~~~ & \tt union~\_uex~\{\\
\tt ~~~~~~~~~~~~~~~~~~~~~~~~~~a:integer\_32, & \tt ~~~~~~~~int~a;\\
\tt ~~~~~~~~~~~~~~~~~~~~~~~~~~b:integer, & \tt ~~~~~~~~long~b;\\
\tt ~~~~~~~~~~~~~~~~~~~~~~~~~~c:float & \tt ~~~~~~~~double~c;\\
\tt ~~~~~~~~~~~~~~~~~~~~~]). & \tt \};\\ \hline
\multicolumn{2}{l}{\tt make\_size(Width,~Height,~SizeStr)~:-}\\
\multicolumn{2}{l}{\tt ~~~~new(size,~SizeStr),}\\
\multicolumn{2}{l}{\tt ~~~~put\_contents(SizeStr,~width,~Width),}\\
\multicolumn{2}{l}{\tt ~~~~put\_contents(SizeStr,~height,~Height).}\\
\end{tabular}

}
\caplab{fig:strex}{Left: a \emph{foreign\_type} declaration, a feature of
  the \texttt{structs} module.
  Right: The corresponding, automatically generated C header file
  containing type declarations.
  Bottom: a predicate that creates a \texttt{size} struct with given
  \textit{Height} and \textit{Width}.}
\end{figure*}

\paragraph{C-to-Prolog interface.}  This interface provides services to
start a query to a Prolog goal, request the next solution to a query,
commit to the current solution of a query, and close a query.
Exceptions can be raised in Prolog and inspected in C.  Type check and
conversion functions from Prolog to C and vice versa are available.  C
code accesses Prolog terms only via \emph{SP\_term\_refs}, which are
handles under the control of the memory manager, so that e.g.\ the
garbage collector can function correctly with this interface.  The
C-to-Prolog and Prolog-to-C interfaces are re-entrant to arbitrary
depth.

\cbstart
\subsection{Source-linked debugging}
\cbend

The ability to step through program execution with the current line of
code being highlighted is a crucial piece of debugger functionality, witness
e.g.\ \texttt{gdb} for C, and Prolog is no exception.  This functionality
was designed and implemented for \SP\ around 1997 by Péter Szeredi.
Using the same infrastructure, when an error exception is raised,
\SP\ tries to precisely pinpoint the responsible line of code.  To
support this functionality, an essential service is a way to read a
Prolog term so that every subterm gets annotated with the line number
on which it occurs.  Another essential service is a data structure
that can map a program location to a filename and a line number.  We
use one mechanism for interpreted code and another one for compiled
(native or virtual) code.

\paragraph{Interpreted code.}  Having read a clause annotated as
mentioned above, the clause is first asserted, obtaining a unique
db\_reference. We then create a \emph{layout table} associated with
this db\_reference and store the filename in it. Treating the
annotated clause as a tree, every path from its root to a leaf or
internal node is stored in the layout table, together with its line
number. A path is simply a list of numbers, e.g., $[3,1,2]$ means
``take the $3^{rd}$ argument of the $1^{st}$ argument of the $2^{nd}$
argument of the body''. A custom compressed format is used so as to
minimize space.

During execution of an interpreted clause, it maintains a
virtual program counter, consisting of the db\_reference of the
clause plus the path to the current goal.  This can be maintained very
cheaply.  To identify the line of code in the source, we just look up
the associated layout table, retrieve the filename, and map the path
to a line number.

\paragraph{Compiled code.}  For compiled code, we use a global B-tree
that maps call sites to filenames and line numbers.  Having read a
clause annotated as mentioned above, the line number information is
threaded through the compiler to its back-end, which actually stores
the virtual code in memory.  When the back-end is about to store a
\texttt{call} or \texttt{execute} instruction, it adds the call site
and associated filename and line number to the B-tree.

The VM emulator has a register holding the most recent
call site.  During tracing of compiled code, the emulator escapes to 
an entry-point of the debugger, passing the value of this register.
Using the value, the associated filename and line number are looked up
in the B-tree.

\cbstart
\subsection{Operating system interface}
\cbend

Interfacing with the underlying O/S and with the file system is
inherently a low level activity. There are a lot of platform specific
details and many operations that can report permanent or temporary
failures. In addition, every O/S to which \SP\ has been ported has
idiosyncrasies, like operations that do not work for all types of
streams, or for streams but not process handles, or vice versa.

Prolog programming, on the other hand, is a high level activity and we
want to hide as much as possible of the underlying complexity and
provide an interface to the O/S that ``just works'' and is portable
across major platforms such as UNIX and Windows as well as to more
exotic platforms where \SP\ is sometimes used, such as mobile phones.

\SP~3 interfaced to the O/S using the mechanism provided by the
standard \texttt{stdio} library and its I/O operations. This design
made sense at a time when characters were 7-bit ASCII, Microsoft
Windows was irrelevant, threads did not exist, and (standardized) UNIX
was not widely adopted.  This lowest common denominator strategy eased
portability but also severely limited the features that could be
offered to the Prolog programmer.

With \SP~4 we took the opportunity to redesign the interface to the
underlying O/S and its I/O operations in a way that directly uses the
native capabilities of the underlying O/S. This new interface was code
named the \emph{\SP\ I/O library} (SPIO).

\paragraph{Non-blocking and interruptible operations.}
Some operations, especially I/O related, can take a long time or even
block indefinitely.  In threaded languages, like Java, it is common to
handle this by simply spawning a new worker thread that handles the
blocking operation, while the main program can either wait for the
spawned thread to complete, or can continue to run while the operation
completes in the worker thread.  Non-blocking and interruptable
operations are crucial for multiple reasons:

\begin{itemize}
\item During development, the programmer must be able to interrupt a
  debugged program without terminating the process or otherwise
  corrupting its state.
\item Server applications that need to keep responding to clients
  while at the same time performing I/O.  They must be able to wait for
  either of several I/O operations to complete.
\item \SP\ has a feature called \emph{asynchronous events}.  Such events can
  be posted from C by an arbitrary thread of the process and will
  cause some associated procedure, which can call Prolog, to be called
  by the Prolog main thread. When such an event is posted, any
  blocking I/O must be interrupted so that the event can be
  processed. Internally, asynchronous events are used for signal
  handling, the timeout facility etc.
\end{itemize}

The standard C library provides no non-blocking operations and no way
to wait for I/O to complete. In \SP~3, some low level routines were
used together with \texttt{stdio} streams to provide waitable
I/O. However, mixing \texttt{stdio} stream operations and O/S-level
stream operations does not always work well or even correctly and does
not work at all for some types of streams.

\SP~4 does not use \texttt{stdio} for I/O. Instead, the use of native
O/S routines allows us to wait on, and to do non-blocking I/O to, many
kinds of O/S streams. Unfortunately, not all streams can be handled in
this way. In fact, neither UNIX nor Windows provides non-blocking
primitives that works for all, or even for most, I/O
operations. Instead, SPIO uses worker threads in C, when needed, to
provide the appearance of non-blocking and interruptible blocking
operations.  SPIO also provides the necessary operations for symbolic
streams that do not use an underlying O/S stream, e.g.\ streams used
for reading from a string. Thus, in Prolog code, and code
that uses our C API, the high level I/O functionality ``just works'',
regardless of the type of stream.

The availability of non-blocking streams makes it possible to wait for
multiple streams to become readable or writable, thus enabling server
applications to be written in Prolog. It also allows a debugged
Prolog program to be interrupted, even if it is waiting for I/O to
complete, without disturbing the I/O operation.

\paragraph{File system.} 
File names with non-ASCII characters are handled differently by
different operating and file systems. SPIO
ensures that such file names behave correctly on systems like Mac~OS~X
and Windows, which use Unicode file names. The standard UNIX way of
handling file name encoding, based on a process-specific \emph{locale}, is
arguably broken by design and is largely ignored by SPIO. Instead, SPIO
falls back on UTF-8 on such systems. SPIO permits file names and file
paths to be as long as the underlying O/S can handle. Thus,
the Prolog programmer is not restricted by the limited
length supported by \texttt{stdio}.

\paragraph{Processes.}  SPIO handles all command line quoting and
argument encoding necessary to launch processes on any supported
O/S. SPIO also provides a common abstraction for process handles.  The
Prolog programmer does not need to care about its details, e.g.\ when
passing a non-ASCII file name, with embedded spaces, as an argument to
a launched program and then waiting for the sub-process to terminate.

\paragraph{Unicode and character encodings. } A number of character
encodings are provided for encoding and decoding file and stream
contents. In many cases, SPIO can automatically detect the encoding
used when reading data from a file.

Non-trivial character sets such as Unicode, and non-trivial encodings
such as UTF-8, place special requirements on the implementation. For
instance, it is possible to get an error when writing a character code
that cannot be represented in the encoding used by the stream being
written to.  Such write errors raise an I/O exception.  Similarly,
an exception is raised if the file contains byte sequences that are
invalid in the given encoding.


\cbstart
\subsection{Attributed variables and constraint solvers}\label{sec:attv}

\SP\ was possibly the first Prolog implementation to incorporate
Holzbaur's seminal idea about attributed variables as a way to extend
unification~\cite{PLILP92}.  Attributed variables are involved in two
related mechanisms: (i) suspending a goal on a variable, i.e.\ until
that variable has been bound, and (ii) a means of associating data
with a variable while that variable is not yet bound.  The first
mechanism is implemented by the \texttt{freeze/1}
predicate~\cite{ICLP87a} together with the generic overflow mechanism:
binding the variable will set the generic overflow flag, and running
the suspended goal will be handled by the generic overflow handler, as
described earlier.

The second mechanism allows Prolog code to refer to attributes by names which are
declared per module.  Once the attributes have been declared,
attribute values can be attached to, modified, and detached from any
variable.  On backtracking, such changes are undone.  A module that
has declared some attributes may also define several local ``hook''
predicates, which add extra functionality, needed by
constraint solvers in particular.  The most important such predicate is
\texttt{verify\_attributes(\textit{AVar},\textit{Value},\textit{Goals})},
which extends default unification as follows.  The predicate is called
by the generic overflow handler
whenever a variable \textit{AVar} with attributes in the given module
is about to be bound to a non-variable term or another attributed
variable \textit{Value}.  It is expected to return in \textit{Goals} a
list of goals.  The suspended unification resumes after the call to
\texttt{verify\_attributes/3}.  Finally, the goals in \textit{Goals}
are called.

Figure~\ref{fig:domv} shows the internal representation of attributed
variables, as used by the CLPFD solver.  References to
attributes by name in the Prolog code are translated by macro
expansion to more direct accesses into this representation.  When
attribute values are attached, modified or detached, destructive
updates are used if they are safe.  Otherwise, the internal
representation is partly copied, and the value cell is bound to the
copy.  Once the value cell has been bound, the extra data structures
are no longer reachable and so are subject to normal garbage
collection.

Attributed variables are a crucial mechanism for constraint solvers in
at least B, Ciao, ECLiPSe, GNU, SICStus, SWI and Yap Prologs.  \SP\
has constraint solvers over Booleans~\cite{TR91:09}, rationals and
reals~\cite{CLPQR95}, finite domains~\cite{PLILP97} and
CHR~\cite{CHR04}.

The finite domain solver has grown into a significant subsystem,
comprising some 60,000 lines of C and 9,500 lines of Prolog code.  The
code is dominated by implementations of propagators for global
constraints.  Two attributes are used for a given domain variable $x$,
as shown in Figure~\ref{fig:domv}.
Constraint propagation is driven by domain changes as
opposed to variable bindings, and so the solver uses its own
propagation loop instead of the \texttt{freeze/1} mechanism.
The solver resides in the \texttt{clpfd} Prolog module, which also
exploits some extensions to the Prolog system:

\begin{description}
\item[New predicate type.]  So-called indexical
  propagators~\cite{PVH91} for smallish constraints can be expressed
  in a special stack machine language.  The solver provides a compiler
  into this language as well as an ``assembly code'' notation.  Such
  propagators are seen by Prolog as predicates of a specific
  type---the constraint is posted simply by calling the predicate.
  Whenever the VM emulator encounters such a call, it escapes to
  \texttt{clpfd:solve/2}, the relevant solver entrypoint.  The binary
  file format also needed to be extended to accommodate these
  predicates.

\item[Global term references.]  The global constraint propagators are
  stateful.  They maintain the constraint arguments as well as
  auxiliary data structures in a block of memory.  This requires a way
  to store a persistent reference to a Prolog term in a C variable.
  The SP\_term\_ref mechanism mentioned earlier is however not
  persistent---an SP\_term\_ref becomes invalid as soon as control
  returns from C to Prolog.  So a persistent variant of term
  references needed to be introduced.

\item[Memory management.]  The solver C code has a license to
  penetrate the normal memory barriers, i.e.\ it can directly
  manipulate the internal term representation, bypassing the normal
  interface functions.  In addition to global term references, the
  solver has other data structures that the Prolog memory manager
  needs to be aware of.  Thus when e.g.\ a heap overflow occurs, the
  memory manager calls certain \texttt{clpfd} interface functions to
  ensure that the solver data structures are processed as need be.
\end{description}
\cbend

\begin{figure*}[tbp]\centering
\cbstart
\begin{tabular}{c}
\framebox[\textwidth][c]{\includegraphics[width=\textwidth]{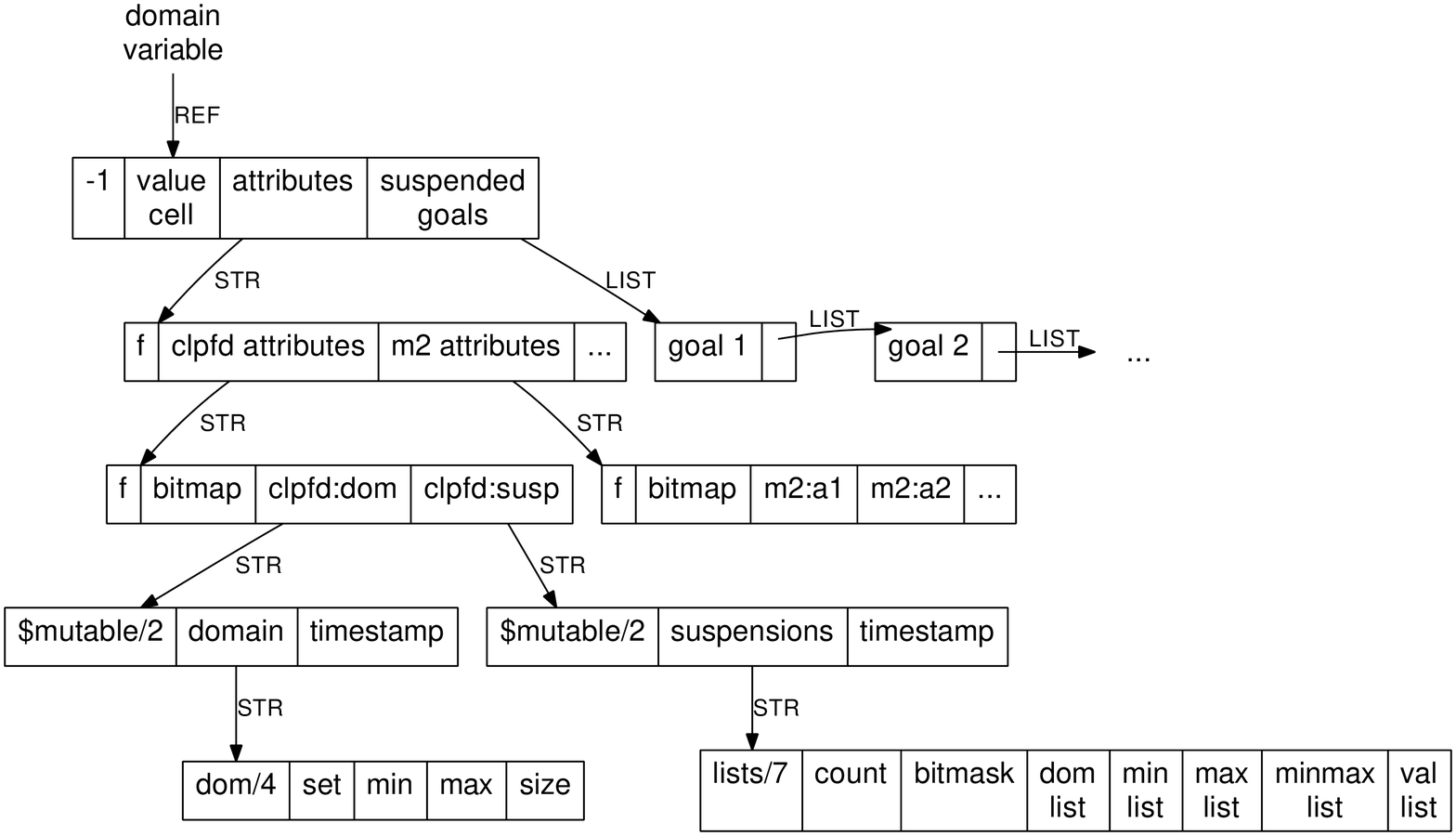}} \\ 
\end{tabular}
\caplab{fig:domv}{Internal representation of domain variables, as a
  special case of attributed variables.
  The root is a reference to a value cell extended with an attributes
  slot and a suspended goals slot.  The value cell is a self-reference
  while the variable is unbound.  \SP~3 used a dedicated tag for
  attributed variables, represented as three consecutive words (value
  cell, attribute slot, suspension list).  \SP~4 uses a generic
  variable tag, but the three words are preceded by a word containing
  -1.  Together with an address comparison, this suffices to
  distinguish attributed variables from normal variables.  This
  distinction needs to be made mainly when a variable is bound: if it
  is attributed, the generic overflow flag is set.  The attributes
  slot contains a structure with one component per module
  (\texttt{m1}, \texttt{m2}, \ldots) that has declared
  attributes. Each such component is a structure with the actual
  attribute values, plus a bitmap indicating whether or not each given
  value is present. The suspended goals slot contains a plain list of
  goals, i.e.\ the \texttt{freeze/1} mechanism can suspend more than
  one goal on the same variable.  The CLFPD solver uses two attributes, both holding
  a mutable, for a given domain variable $x$. \texttt{dom/4}
  stores its domain, while \texttt{lists/7} encodes the dependency lists,
  i.e.\ the set of constraints mentioning $x$ as well as what kind of
  domain change should schedule each given constraint.}
\cbend
\end{figure*}

\subsection{Miscellaneous}

\SP\ uses a large number of implementation techniques that are shared
with other implementations, Prolog or otherwise.  Some of these
features can be traced back to a source; others are folklore.  We now
list a few of these points.

\begin{description}
\item[Cyclic term unifier.]  Without special care, the unification
  algorithm may not terminate on cyclic terms.  In~\cite{Colmer82}, a
  simple method to avoid this problem is described. Briefly, before
  recursively unifying the $i^\mathrm{th}$ argument of two compound
  terms $p$ and $q$, the unifier temporarily sets the memory cell
  holding $p[i]$ to $q[i]$ (or vice versa). If the unifier later
  encounters the same pair of memory cells, it will see two identical
  terms instead of falling into infinite recursion. Before returning,
  the unifier restores all such modifications.  We use the same
  technique in the term comparison algorithm that determines the
  relation between two given terms in the standard order of terms.

\item[Mutable terms.]  
  \begin{sloppypar}
  \SP\ used to
  have a non-logical feature called \texttt{setarg($I$,$P$,$X$)}.  The
  effect is to set the $I^\mathrm{th}$ argument of the compound term $P$ to
  $X$, restoring the old value on backtracking.  To support restoring,
  the trail must be generalized to accommodate such old values and
  their destinations.  This feature exists in at least
  B, Bin, Ciao, ECLiPSe, GNU, SWI, and Yap Prologs. 
  Around 1995, we replaced
  \texttt{setarg/3} by a new abstract datatype \emph{mutable
  term} with operations to create such a term and to get and update
  its value.  The implementation is based on~\cite{SPLT90}: each
  mutable term has a time-stamp, which indicates when the value
  was last updated.  The point is, if no choicepoint has been pushed
  between two updates, the second update does not need to be trailed.
  We also extended the variable shunting algorithm~\cite{Sahlin91} to 
  compress reset chains for mutables.  We treat mutable terms as
  non-ground, 
  no matter what the current value is.  Subsequently,
  mutable terms have been adopted by Yap Prolog.
  \end{sloppypar}

\item[Bignums.]  Bignums are available in at least Ciao,
  ECLiPSe, SICStus, SWI, and Yap Prologs.
  We do not use any publicly available
  multiprecision libraries, since when our code was developed, none of
  the available libraries was compatible with our particular memory
  management requirements.

\item[Asserting clauses and copying terms.] Internally, these two operations
  are very similar and share much of the code.  Both use variants of
  Cheney's algorithm~\cite{CACM70}.  The main difference is in the
  output: the assert operation creates an interpreted clause, i.e.\ a
  kind of blue-print from which a brand new clause copy can be built in
  linear time, whereas the copy operation creates a new term directly.

\cbdelete
\item[Object-oriented programming.]  Although the combination of logic
  programming and object-oriented programming was never a research
  topic at SICS, \SP\ does provide such modules.  The \SP~3
  \texttt{objects} module was designed with an emphasis on knowledge
  representation.  
\cbstart
  It was based on the notions of prototypes,
  inheritance and delegation.  The implementation piggybacked on the
  module system: a named object was represented by the Prolog module
  with the same name, resulting in an obvious risk for name clashes.
  Furthermore, the module data structures and primitives had to be
  extended in order to provide all the services that the object system
  needed. 
\cbend

  The \SP~4
  \texttt{objects} module is based on the notions of classes and
  inheritance.  The emphasis is on efficiency.  The
  implementation is 100\% based on source-to-source compilation and
  does not rely on or extend the module system.  A detailed
  description can be found in~\cite{CSAI11}.

\item[Exceptions, or \texttt{catch} and \texttt{throw}.]
  We use the implementation proposed in~\cite{CW96}.

\item[Cleaning up, or \texttt{call\_cleanup}.] A very
  common situation in programming is the following.  Some algorithm
  needs to run, holding some resources.  Those resources must be freed
  afterwards, no matter whether or not the algorithm terminates
  normally.  Common Lisp provides a primitive for this purpose:

  \begin{tabular}{c}
  \tt (unwind-protect \textit{protected} \textit{cleanup}) \\
  \end{tabular}

  \noindent
  which evaluates the form \textit{protected} in a context where
  the form \textit{cleanup} is guaranteed to be executed when and if
  control leaves the form \textit{protected} by \emph{any} means.
  Finally, the value of \textit{protected} is returned from the
  \texttt{unwind-protect} form.  

  Around 1997, the first author introduced an analogous construct into
  \SP, naming it \texttt{call(\textit{Goal},\textit{Cleanup})}.
  Richard O'Keefe criticized him for this choice of name, which
  clashes with the multiple argument generalization of
  \texttt{call/1}.  Richard was absolutely right of course, and the
  construct was later renamed to \texttt{call\_cleanup/2}, its present
  name.  Subsequently it has found its way into at least B, ECLiPSe,
  SWI, XSB, and Yap Prologs.

  \texttt{call\_cleanup/2} guarantees the
  execution of \textit{Cleanup} if \textit{Goal} succeeds
  determinately, fails, or raises an exception.  Also, if
  \textit{Goal} succeeds with some alternatives outstanding, and those
  alternatives are removed by a cut or an exception, \textit{Cleanup}
  is executed.  The implementation is composed of the following elements:
  \begin{itemize}
  \item \textit{Cleanup} goals are placed on the trail. The general
    backtracking mechanism simply executes such goals as they are
    encountered on failure or exception.
  \item A bit $c(b)$ is reserved in every choicepoint $b$, denoting the fact that
    there may be a pending \textit{Cleanup} goal when $b$ equals
    the current choicepoint $B$.
  \item When \texttt{call\_cleanup} is called, $b_0 \gets B$, $c(b_0)$ is set,
    and \textit{Cleanup} is pushed on the trail.
  \item On non-deterministic exit from \texttt{call\_cleanup}, $c(b)$ is set for
    all choicepoints $b$ that predate $b_0$, so as to ensure that
    \textit{Cleanup} is run if and when a cut back to $b_0$ or beyond occurs.
  \item On deterministic exit from \texttt{call\_cleanup}, and upon execution
    of a cut, if $c(B)$ is set, the generic overflow flag is set.
  \item If the generic overflow handler finds a cleanup goal in the
    current trail segment, it arranges for it to be run.  It clears
    $c(B)$ if appropriate.
  \end{itemize}
\end{description}








\cbstart
\section{Development environment}
\label{sec:env}
\cbend

\subsection{Background}

Since early on, \SP\ has had an Emacs-based development environment,
with syntax highlighting, source-linked debugging, links to the
documentation, and more.  However, both our Emacs-based development
environment and Emacs itself lacks many of the features that users
have come to expect from a modern integrated development environment
(IDE), such as:

\paragraph{Parser.} 
Anything but the most trivial language support requires a proper
parser, including support for operator directives. Without a parser it
is not possible to get much more advanced than showing variables in
italics. The parser must be part of the IDE, 
\cbstart
as running it in a separate process
would likely cause intolerable response times.
\cbend

\paragraph{Semantic analysis.}
The dynamic nature of Prolog is an advantage for the developer but
makes it difficult for the compiler to provide
diagnostics. Traditionally, like most other Prolog
implementations, \SP\ warns about syntax errors but provides little in terms
of semantic diagnostics.  Semantic diagnostics are mostly limited
to local issues such as singleton variables and discontiguous
clauses.
While \SP\ comes with several useful tools that provide more
advanced diagnostics, e.g.\ for determinacy checking and cross
referencing, these tools must be run separately, which is inconvenient.
On the other hand, an IDE, especially if it has knowledge about the
set of files that makes up a Prolog program, can provide the same and more
functionality than the existing tools, while the user
edits or browses the program files. An IDE can also give feedback from
syntactic and semantic analysis in a more useful way than what is
possible with separate tools, e.g.\ by highlighting undefined predicate
calls or incorrect predicate arguments directly in the source code
editor.

\paragraph{Code refactoring.}
Code refactoring means automatic and usually global changes to a
program, preserving the semantics of the program. Typical examples for
Prolog are: renaming a predicate, reordering the arguments of a
predicate, or adding arguments to a predicate, automatically updating
all callers.

\paragraph{Scalability.}
Our commercial customers have applications comprising hundreds of
modules adding up to several hundred thousand lines of code.  This
fact stresses the importance that our IDE be scalable to such code sizes.


\paragraph{Implementation.}
We have implemented our IDE in Eclipse, an application and IDE
framework written in Java. Eclipse has already proved itself as a
foundation for powerful IDEs for many programming languages. Using
Eclipse as the basis for an IDE also gives many features for free,
such as portability, integration with common revision control systems
and support for multiple programming languages in the same IDE. Using
Eclipse will also make it possible to integrate other tools such as
profiler and constraint visualizers into the IDE. In addition, Eclipse
makes it possible for us to package our IDE as a standalone product
with a completely Prolog-centric appearance, if needed.  

A first version of the IDE, with working name SPIDER, was released
together with \SP~4.1, in December 2009. It is still in beta and lacks
some of the planned features but it is already quite useful and its
analysis functionality has helped us identify and fix several defects
in our own code.

\cbstart
\subsection{SPIDER in action}
\cbend

Figure~\ref{fig:spider} shows some of the features of SPIDER in action.
We now discuss some of its central features:

\begin{figure*}[tbp]\centering
\begin{tabular}{c}
\mbox{\includegraphics[width=\textwidth]{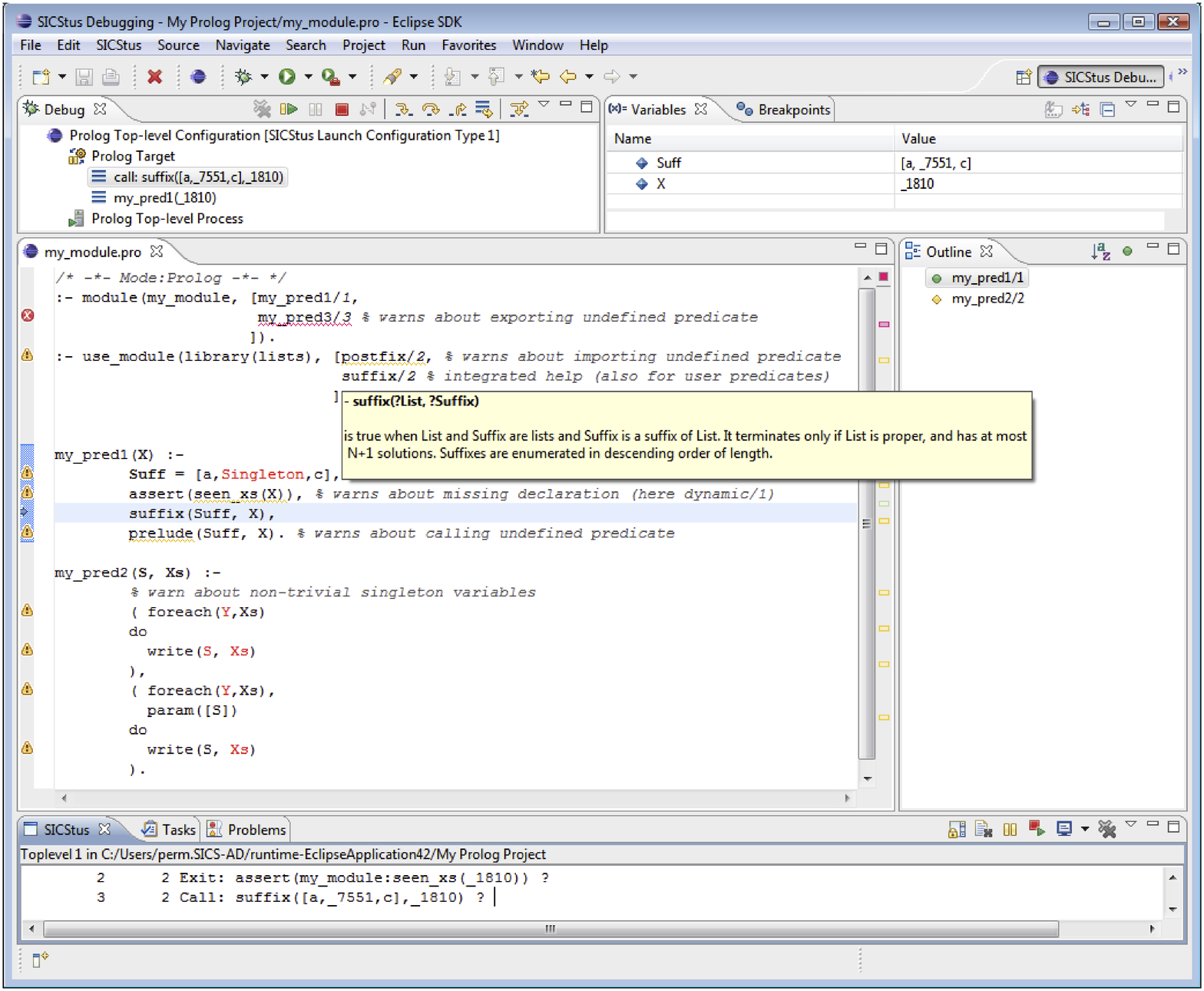}} \\ 
\end{tabular}
\caplab{fig:spider}{\SP\ IDE window.
Top left: debugger pane.
Top right: variable pane.
Middle left: source code pane with highlighting and pop-ups.
Middle right: outline pane.
Bottom: toplevel pane.}
\end{figure*}

\paragraph{Editor.}
While editing, SPIDER continuously re-parses the code and annotates
the text with warnings and semantic highlightings.  Warnings include:
calls to undefined predicates, import or export of predicates not
defined in module, assert of predicate not declared dynamic, not using
\texttt{use\_module/[1,2]} when loading a module file, singleton
variables.

Semantic highlightings include a special appearance of first and
single occurrences of a variable.  This is done also in the context of
disjunction and logical loops~\cite{Schimpf02}, when the variable may
have more than one \emph{semantically} first or singleton occurrence.

Calls to undefined predicates are highlighted, including when they
appear as arguments to meta predicates.

The editor provides completion of predicate names and documentation
pop-up when the mouse pointer is hovering over a predicate name. The
documentation is formatted on-the-fly for user written code and there
is an integrated browser for the \SP\ product documentation.

The definition of a user-defined or built-in library predicate or
module can be opened with a single click or keyboard command.

\paragraph{Toplevel.}
The toplevel implements the traditional terminal interface and
provides a familiar interface, including the traditional debugger.

\paragraph{Debugger.}
The debugger shows an ancestor stack, local variable bindings and
direct access to some common debugger control commands, like step
into, step over and redo. The traditional terminal-based debugger
interface is active at all times in the toplevel, so the power user is
free to use that, if desired.

The debugger and editor together provide a point and click interface
for setting line breakpoints and spypoints. It is also possible to
temporarily disable all breakpoints and to save breakpoints
across debugging sessions.

The debugger and toplevel can attach to a running \SP\ process that may be
running on another machine (and platform) than the IDE. This is useful
for those that embed \SP\ as part of a larger program or system.

\paragraph{Future features.}
A prerequisite of many types of program analysis is complete
information about all source code in a program. This requires not only
knowing which files make up the program but also how these files load
each other, especially when modules are distributed among multiple
non-module files. SPIDER, like many other Eclipse-based language
environments, delegates this task to a separate \emph{indexer}, which
updates the information as files are modified.  The indexer
functionality of SPIDER is currently being implemented. When this work
is completed, we plan to add features such as call hierarchy and
determinacy analysis, providing similar functionality as that of our
current \texttt{spxref} and \texttt{spdet} tools, but with immediate
feedback as the program is modified. The indexer is also a
requirement for refactoring and other planned features that currently
have no counterpart among the existing \SP\ tools.

\section{Applications}
\label{sec:app}

\SP\ is being used on a 24/7 basis in major applications comprising
hundreds of modules adding up to several hundred thousand lines of
code.  It is a pity, but for reasons of customer confidentiality, we
are not at liberty to describe some of the most impressive
ones. Anyway, we now briefly describe some applications for which
permission has been generously granted, or where the information is
publicly available.

\paragraph{Speech recognition.} 
Clarissa\footnote{\url{http://ti.arc.nasa.gov/project/clarissa/}},
a fully voice-operated procedure browser was developed
by the NASA Intelligent Systems Division.
On the International Space Station (ISS), astronauts execute thousands
of  complex  procedures  to  maintain  life support systems, check out
space  suits  and conduct science experiments, among their many tasks.
Today,  when carrying out these procedures, an astronaut usually reads
from  a  PDF viewer on a laptop computer, which requires them to shift
attention  from  the task to scroll pages. Clarissa enables astronauts
to be more efficient and to give full attention to the task while they
navigate through complex procedures using spoken commands.

Clarissa was implemented mainly using \SP\ and a speech
recognition     toolkit    provided    by    Nuance    Communications.
Application-specific  spoken  command  grammars were constructed using
the \SP\ based Regulus platform~\cite{Regulus}.

\paragraph{Telecom.}
Ericsson Network Resource Manager (NRM) provides the capabilities for
configuring and managing complex multi vendor IP Backbone
networks. NRM assists the operator in making decisions when planning,
configuring and making configuration changes.

The modeling part of the NRM software, an expert tool assisting the
network operator, was implemented in \SP. The constructed
network model, created by analyzing the actual router configurations,
is used both for showing a graphical representation and for
validating the network.

\paragraph{Biotech.}
A dispensation order generation algorithm for
Pyrosequencing's sequence analysis instruments, using 
constraint programming in \SP~\cite{APBC2004,CP2004:pyro}.
The algorithm can be described as a compiler, which calculates an
instruction sequence based on an input specification.
Applications include genetics, drug discovery, microbiology, SNP and
mutation analysis, forensic identification using mtDNA,
pharmacogenomics, and bacterial and viral typing.

\paragraph{Logistics.}
One of the products of
RedPrairie Corporation, a leading provider of real-time
logistics solutions, is a real-time optimization engine,
COPLEX.  The kernel of the engine is written in
\SP\ using its finite domain constraint solver library.

\paragraph{Data Mining.}
Compumine AB's data mining software Rule Discovery System (\RDS)
is a tool for generation of reliable, accurate, and interpretable rule
based prediction models by automatically searching databases for
significant patterns and relationships. \RDS\ was
implemented in \SP\ and has been successfully applied to problems in a
large number of data intensive areas such as pharmaceutical research,
language technology, and engineering. 

\paragraph{Business Rules: The $360^o$ Fares System.}
The paper~\cite{Wilson05} describes an application
running the $360^o$ Fares System.  It is one of the largest and most profitable
Prolog applications written. Prolog is the business-rule component in
a multi-component application that includes network, user interface,
and security data access tiers.

\paragraph{Biomedical text search.}
MetaMap\footnote{\url{http://metamap.nlm.nih.gov/}}
was developed by Alan Aronson at the National Library of
Medicine (NLM) to map biomedical text to the UMLS Metathesaurus or,
equivalently, to discover Metathesaurus concepts referred to in
text. MetaMap uses a knowledge intensive approach based on symbolic,
natural language processing (NLP) and computational linguistic
techniques. 
MetaMap is one of the foundations of NLM's Medical Text
Indexer (MTI) which is being applied to both semiautomatic and fully
automatic indexing of biomedical literature at NLM.  MetaMap was first
implemented in \QP\ and is being ported to \SP.

\paragraph{Safety-critical applications.}
\cbstart
SPARK\footnote{\url{http://www.praxis-his.com/spark.aspx}}~\cite{Barnes2003} 
\cbend
is a high level programming language and toolset designed for
writing software for high integrity applications.  SPARK enables the
application of formal verification techniques in a segregated monitor
architecture, ensuring rapid compliance.  The SPARK toolset comes in a
GPL version and includes a theorem prover implemented in \SP.

\section{Conclusion}
\label{sec:con}

Now that the system has been around for nearly 25 years, a relevant
question to ask is: what are the key good and less good design
decisions?  We now try to give some answers.

First of all, there hardly were any truly bad decisions.  Some
decisions, like endeavoring into compiling to native code, meant huge
amounts of work for platforms that eventually went extinct.  But at
the same time, good research was done, important lessons were learned,
and pieces of technology were developed that could be reused in a JIT
compiler, for example.

\begin{sloppypar}
One questionable decision was the fact that \SP~3 supported two
dialects, ``classic'' and ISO, in the same system, and even let the
user dynamically switch between the two.  This made it awkward to
document certain built-in predicates, like \texttt{atom\_chars/2},
whose semantics differs from dialect to dialect, as well as all the
other, subtler differences.  It also made it quite a challenge to
ensure that all library modules would run in both dialects.  We are
not aware of any other programming system, Prolog or otherwise, that
provides this degree of freedom.  Of course, this situation stemmed
from the fact that the ISO standard was published quite late, when a
lot of application code had already been written by users as well as
implementers.  We wanted to promote the ISO standard, but at the same
time, we had no right to break people's existing code.  Our solution
to this dilemma was a dual dialect system.
\end{sloppypar}

A lesson that keeps getting reiterated is the importance of backward
compatibility.  For obvious reasons, users are very unforgiving to
changes in behavior of the programming system, even if it concerns
minor points that are not necessarily specified in detail in the
documentation.  For example, at one time we were flamed by a customer
for changing the printed appearance of certain floating-point numbers,
although the old and new appearances were both legal syntax.  There is
no escape from this issue, and the Prolog standardization committee is
well advised to bear it in mind.  The first author knows from first
hand experience as a committee member how tempting it is to
start ``cleaning up'' or ``redesigning'' parts of the language.  Such
ambitions can be commendable, but at this stage they are only viable if full
backward compatibility can be preserved.

Finally, the quality of the POSIX primitives \texttt{malloc()} and
\texttt{free()} in today's operating systems is probably high enough
to make a dedicated memory manager redundant.  However, we do have
customers that depend on the ability to control memory allocation with
tunables, and it is not clear whether their applications would run with
tolerable performance without a tunable, dedicated memory manager.

But by and large, \textit{je ne regrette rien}.

\cbstart
\section*{Acknowledgment}
\cbend

A large number of people have contributed to the development of
SICStus Prolog, including:
Jonas Almgren,
Johan Andersson,
Stefan Andersson,
Nicolas Beldiceanu,
Tamás Benk\H{o},
Kent Boortz, 
Per Brand,
Göran Båge, 
Per Danielsson, 
Joakim Eriksson, 
Jesper Eskilson, 
Niklas Finne, 
Lena Flood, 
György Gyaraki,
Dávid Hanák, 
Seif Haridi, 
Ralph Haygood, 
Christian Holzbaur, 
Key Hyckenberg, 
Carl Kesselman,
Péter László, 
Carl Nettelblad,
Greger Ottosson, 
Dan Sahlin, 
Peter Schachte, 
Rob Scott, 
Thomas Sjöland, 
Péter Szeredi, 
Tamás Szeredi, 
Johan Widén, 
Magnus Ågren,
Emil Åström
and the authors.

The Industrialization of \SP\ (1988-1991) was funded by
Ericsson Telecom AB, NobelTech Systems AB, Infologics AB, and
Televerket, under the National Swedish Information Technology Program IT4.
The development of \SP~3 (1991-1995) was funded in part by
Ellemtel Utvecklings AB.

We thank Magnus Ågren and the anonymous referees for their constructive comments.

\bibliographystyle{acmtrans}
\bibliography{TPLP09}
\end{document}